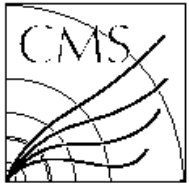 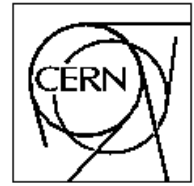

**The Compact Muon Solenoid Experiment**

**CMS Note**

Mailing address: CMS CERN, CH-1211 GENEVA 23, Switzerland

# On a Possibility to Measure the Magnetic Field Inside the CMS Yoke Elements

V. I. Klioukhine[1], R. P. Smith[2]

*FNAL, Batavia, IL, U.S.A.*

B. Curé

*CERN, Geneva, Switzerland*

C. Lesmond

*CEA/Saclay, Gif-sur-Yvette, France*

**Abstract**

A procedure to measure the magnetic field inside the CMS yoke elements is considered. Fast discharge of the CMS coil can be used to induce voltages in flux loops installed around selected elements of the CMS yoke. By sampling the voltage induced in any one loop, and integrating the voltage waveform over the time of the discharge, the total initial flux in the loop can be measured. The average value of the magnetic field in the yoke element normal to the plane of the flux loop is obtained by dividing the measured value of the flux by the known area enclosed by the loop and the number of turns in the loop.

[1] E-mail: klyukhin@fnal.gov

[2] E-mail: rpsmith@fnal.gov

# 1 Introduction

In this note a study is undertaken to determine if values of the magnetic field in the CMS yoke [1] can be measured by the use of flux loops installed around selected elements of the yoke. Voltages will be induced in such flux loops during times of change of the magnetic field of the magnet. Provided the voltages are measurably large and sufficiently noise-free, it is in principle possible to integrate the induced voltages and deduce the total flux change in the steel during the time of field change. The largest voltages are induced during times of the most rapid field change of the magnet, and the voltage induced in any one flux loop is proportional to the number of turns in the loop.

The charge time of the CMS magnet is approximately 5 hours and the normal discharge time is about two hours. The fast discharge time constant is approximately 200 seconds. Evidently measuring the flux loop voltages during coil fast discharge provides the most advantageous opportunity to make the measurements envisioned.

During normal operation of the CMS magnet fast discharge of the superconducting solenoid will normally only be triggered by the detection of some abnormal operating condition which makes it advantageous from a safety point of view to discharge the coil as rapidly as is feasible. During the testing of the magnet system on the surface before it is lowered into the CMS collision hall at PA5, the magnet will be subjected to one or more fast discharges to verify the proper performance of the fast discharge system. These periods of fast discharge provide an opportunity to make the flux loop measurements proposed herein.

The fast discharge (and subsequent quench) of the coil has been modeled elsewhere [2] and the total Ampere-turns in the coil and support cylinder versus time calculated. The model ignores any eddy currents in the steel elements of the magnet yoke.

The calculated values of the Ampere-turns in the coil at nine distinct intervals during the first 300 second of the coil discharge have been used to calculate the flux distributions everywhere in the magnet at these times, using the 3D magnetostatic code TOSCA [3]. In section two of this note the geometry of the CMS yoke is described, the assumed magnetic properties of the steel provided, and the boundary conditions used in the TOSCA calculations presented. In section three the time-dependence of the variations of the flux in selected portions of the CMS yoke are presented, and calculations of the voltages that will be induced in appropriate flux loops installed on the iron pieces are made. In the fourth section the results of the calculations are presented, and a discussion of the conclusions to be made from the study is presented in the forth section.

# 2 Description of the CMS Magnet System Model

## 2.1 Geometry of the TOSCA model

The model of the CMS magnetic system presented in Fig. 1 consists of the superconducting solenoid coil and quench-back cylinder, and one quadrant of the 14 m outer diameter and 20 m long steel yoke.

The physical CMS yoke includes the central barrel yoke surrounding the coil, two end-caps located at each end of the solenoid, and the ferromagnetic parts of the hadronic forward calorimeter arranged downstream of the end-caps. The barrel is subdivided into five rings YB-2 through YB+2, each consisting of three concentric approximately circular layers joined by connecting brackets. An additional fourth layer, the tail catcher (TC), exists only inside the central barrel ring. Each end-cap consists of one small nose disk and four large disks YE1 through YE4 mounted axially along the axis of the solenoid and downstream of the barrel rings.

The TOSCA model of the yoke includes the barrel rings (omitting the connecting brackets), plus the symmetric tail catcher, the steel tube supporting the hadronic end-cap calorimeter, the nose disk and the four end-cap disks. It omits the hadronic forward calorimeter ferromagnetic parts. The dimensions of the yoke elements are described elsewhere [4]. Since the time of that note, the description of the yoke has been modified slightly -- the end-cap disks have been moved an additional 25 mm downstream of the end barrel ring as shown in the CMS drawings of January 4, 2000. These modifications are incorporated into the present model.

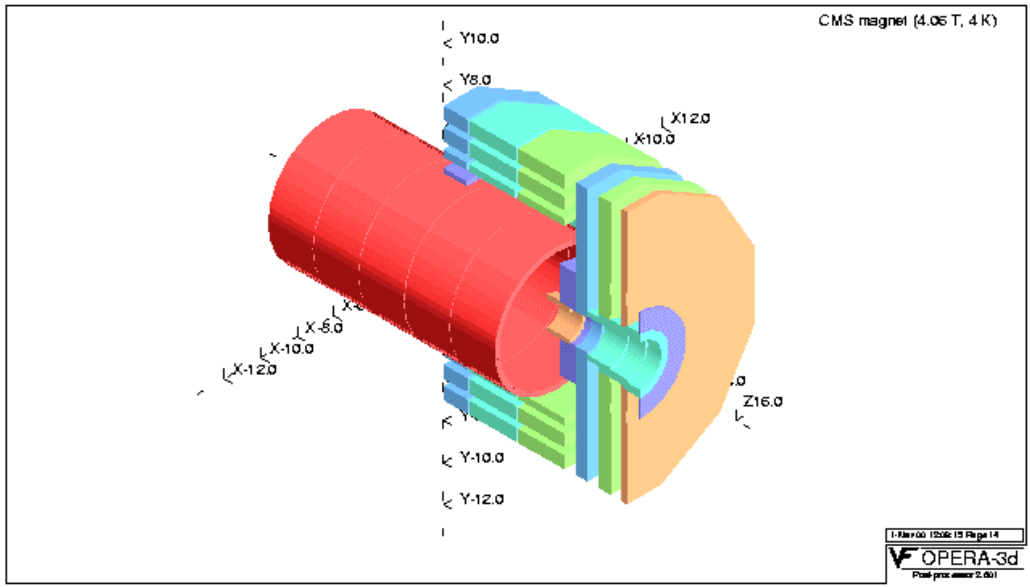

Figure 1: Model of the CMS magnetic system (the quench-back cylinder is not shown).

The finite-element mesh of the model is constructed from 94711 nodes. To describe the model, a Cartesian coordinate system with the origin placed in the center of the solenoid is used. The Z-axis is along the solenoid axis (and pointing out of the plane of the figure) and the Y-axis is vertically upward. The X-axis is in a horizontal plane, completing a right-handed triple.

In the model the coil consists of twenty cylindrical sheets of current deployed concentrically at four distinct radii and in five axial groupings. The radial thickness of each current sheet is 20.63 mm and the radii of the sheets are 3174.93, 3240.06, 3305.19 and 3370.32 mm. The dimensions of the current sheets were chosen to correspond to the positions of the superconducting cable in the windings of the physical coil when the latter is at cryogenic temperatures. The five modular groupings of cylinders are separated axially from one another by 45.86 mm. Each module has a length 2443.76 mm and the overall axial length of the 5-module assembly is 12402.24 mm.

The aluminum quench-back cylinder in the model has a length 12448.12 mm, an inner radius 3413.99 and a radial thickness 49.8 mm. Its dimensions also correspond to that of the physical coil at cryogenic temperatures.

## 2.2 Magnetic properties of steel

To define the magnetic properties of the ferromagnetic parts in the model, we use the B-H dependence presented in Table 1 and used earlier in [4]. Here B, H, and $\mu_{rel}$ denote the magnetic flux density, the field strength, and the relative permeability of the steel, respectively.

Table 1: B-H curve used in the calculations.

| B, T | H, A/m | $\mu_{rel}$ | B, T | H, A/m | $\mu_{rel}$ |
|---|---|---|---|---|---|
| 0.0 | 0.0 | — | 1.8364 | 13089.241 | 111.646 |
| 0.0098 | 17.900000 | 435.676 | 1.8954 | 17579.686 | 85.8031 |
| 0.0139 | 24.718828 | 447.484 | 1.9571 | 23885.627 | 65.2028 |
| 0.0193 | 32.957554 | 466.007 | 2.0000 | 29622.010 | 53.7286 |
| 0.0244 | 40.108746 | 484.106 | 2.0500 | 37078.430 | 43.9970 |
| 0.0283 | 44.799442 | 502.694 | 2.1000 | 47831.691 | 34.9376 |
| 0.0415 | 60.497574 | 545.884 | 2.1500 | 61216.754 | 27.9485 |
| 0.0635 | 81.294975 | 621.584 | 2.1750 | 69701.523 | 24.8317 |
| 0.1065 | 109.68913 | 772.638 | 2.2000 | 79913.414 | 21.9075 |
| 0.2127 | 148.47574 | 1139.99 | 2.2500 | 102183.14 | 17.5224 |
| 0.3882 | 199.97342 | 1544.80 | 2.2800 | 120970.33 | 14.9984 |
| 0.5802 | 270.18683 | 1708.85 | 2.3000 | 136405.58 | 13.4179 |
| 0.7616 | 364.20496 | 1664.07 | 2.3443 | 169844.02 | 10.9838 |
| 0.9266 | 489.62607 | 1505.98 | 2.3996 | 211885.75 | 9.01213 |
| 1.0784 | 659.75281 | 1300.73 | 2.4905 | 282777.06 | 7.00862 |
| 1.2141 | 889.68927 | 1085.94 | 2.5627 | 339680.91 | 6.00367 |

```
1.3275    1191.2357    886.803   2.6706   425818.56   4.99085
1.3600    1278.3292    846.616   2.8498   568454.75   3.98941
1.4249    1495.2793    758.320   3.2074   853096.19   2.99189
1.5013    2179.5076    548.150   3.5644  1137160.0    2.49434
1.5660    2937.9355    424.170   4.2782  1705280.5    1.99644
1.6241    3965.0059    325.956   4.8134  2131080.2    1.79739
1.6799    5350.3267    249.858   5.7052  2840805.8    1.59816
1.7352    7212.6812    191.445   6.4186  3408440.2    1.49856
1.7788    9246.5557    153.087   7.4887  4260027.0    1.39889
```

In Fig. 2 are plotted values of the relative permeability in the CMS yoke at radii 3.93, 4.7575, 5.66, and 6.685 m when the magnet is energized to 4T. These radii correspond to the average radial distances of the plates from the Z-axis. From this figure one can see that in three central rings the plates are almost saturated and the variation of the permeability along the plates is small in this region. In contrast, the variation of the permeability in the outer barrel rings and in the end-cap disks is large.

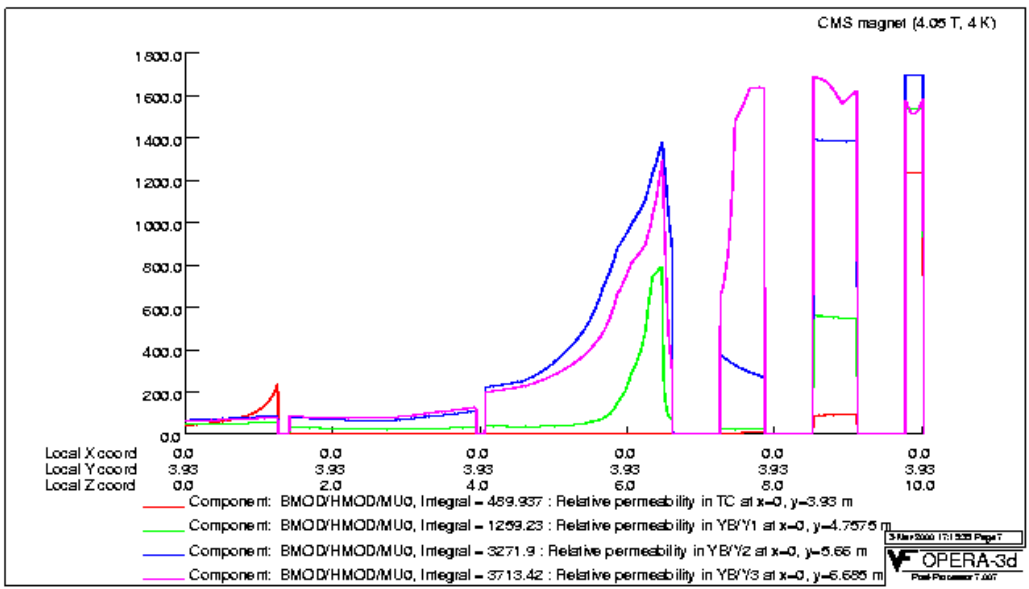

Figure 2: Relative permeability in the yoke at radii of 3.93, 4.7575, 5.66, and 6.685 m.

### 2.3     Boundary conditions for the TOSCA Calculation

The TOSCA code requires the specification of boundary values to define the symmetry chosen for the problem. In the model the normal component of the magnetic flux density is set to zero on the surface of a cylinder with a radius 10 m, and on the plane normal to the axis of the cylinder at $Z = 13.5$ m. The magnetic flux density vector is also constrained to be orthogonal to a plane normal to the axis of the cylinder at $Z = 0$ m.

## 3    Calculation of Magnetic Flux Values during Coil Discharge

The initial total current in the coil, 42.51 MA-turns [5], gives 4.06 T in the center of solenoid. The time dependence of the total current in the winding and quench-back cylinder during coil fast discharge as given in note [2] is shown in Fig. 3.

From the curve of Fig. 3 the values of the total Ampere-turns in the coil and quench-back cylinder at nine different times, 0.0, 50.186, 100.35, 125.117, 151.347, 176.134, 200.546, 251.343, and 305.881 s, were selected for the TOSCA calculations.

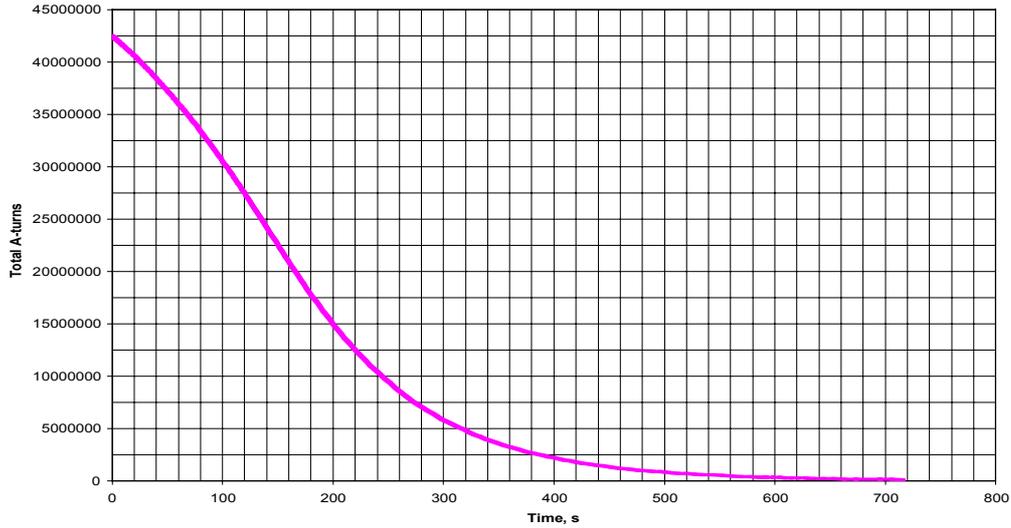

Figure 3: Total current variation during the solenoid fast discharge.

It is of interest to obtain the time-dependence of the magnetic flux in the barrel plates located at the azimuth sectors $\pm 15\,^\circ$ from the Y-axis and in the cross-sections of the first two end-cap disks YE/1 and YE/2 in the azimuth sectors $\pm 10\,^\circ$ from the Y-axis. For this purpose we integrated the calculated magnetic flux densities over the barrel plate and end-cap disk cross-sections placed as shown in Tab. 2.

To find the resulting average flux values, we calculate the cross-sectional areas of the corresponding plates at these same locations. Note that the areas of the barrel plate cross-sections used in the calculations are restricted by the geometry of the connecting brackets. The areas of the cross-sections used in the end-caps are determined by the azimuthal sector of $20\,^\circ$.

Table 2: Disposition of the yoke cross-sections used in the flux calculations.

| Unit | X, m | Y, m | Z, m | Area, m$^2$ |
|---|---|---|---|---|
| TC | −1.0÷0.87 | 3.84÷4.02 | 0 (middle); 0.768 | 0.3366 |
| YB/0/1 | −1.075÷1.185 | 4.61÷4.905 | 0 (middle); 0.768 | 0.6667 |
| YB/1/1 | " | " | 1.918; 2.686 (middle); 3.454 | " |
| YB/2/1 | " | " | 4.574; 5.342 (middle); 6.11 | " |
| YB/0/2 | −1.075÷1.345 | 5.35÷5.97 | 0 (middle); 0.768 | 1.5004 |
| YB/1/2 | " | " | 1.918; 2.686 (middle); 3.454 | " |
| YB/2/2 | " | " | 4.574; 5.342 (middle); 6.11 | " |
| YB/0/3 | −1.655÷1.345 | 6.375÷6.995 | 0 (middle); 0.768 | 1.86 |
| YB/1/3 | " | " | 1.918; 2.686 (middle); 3.454 | " |
| YB/2/3 | " | " | 4.574; 5.342 (middle); 6.11 | " |
| YE/1 | ± 0.4937 | 2.8 | 7.265÷7.865 | 0.59244 |
| " | ± 0.8076 | 4.58 | " | 0.96912 |
| " | ± 1.1814 | 6.7 | " | 1.41768 |
| YE/2 | ± 0.4937 | 2.8 | 8.52÷9.12 | 0.59244 |
| " | ± 0.8287 | 4.7 | " | 0.99444 |
| " | ± 1.1814 | 6.7 | " | 1.41768 |

In the barrel rings YB/0 through YB/2 the cross-sections are orthogonal to the Z-axis and are inset axially 0.5 m from the plate axial ends and in the axial centers of the plates. In the end-cap disks YE/1 and YE/2 the cross-sections are orthogonal to the Y-axis and are placed at 0.17 m, 1.95 (2.07), and 4.07 m radially off the nose disk.

This choice of the cross-section disposition seems reasonable if we look at Figs. 4 and 5 where variations of the average magnetic fluxes through the yoke cross-sections versus Z- and Y-coordinates are presented. From Fig. 4, where the flux of the axial component $B_z$ of the magnetic flux density in the barrel yoke is displayed, one can see that a simple polynomial approximation will adequately describe the axial flux behavior along the barrel plates if the flux is measured in at least three different cross-sections along each plate. The same is valid for the flux of the radial component $B_r$ of the magnetic flux density inside the end-cap segments as shown in Fig. 5.

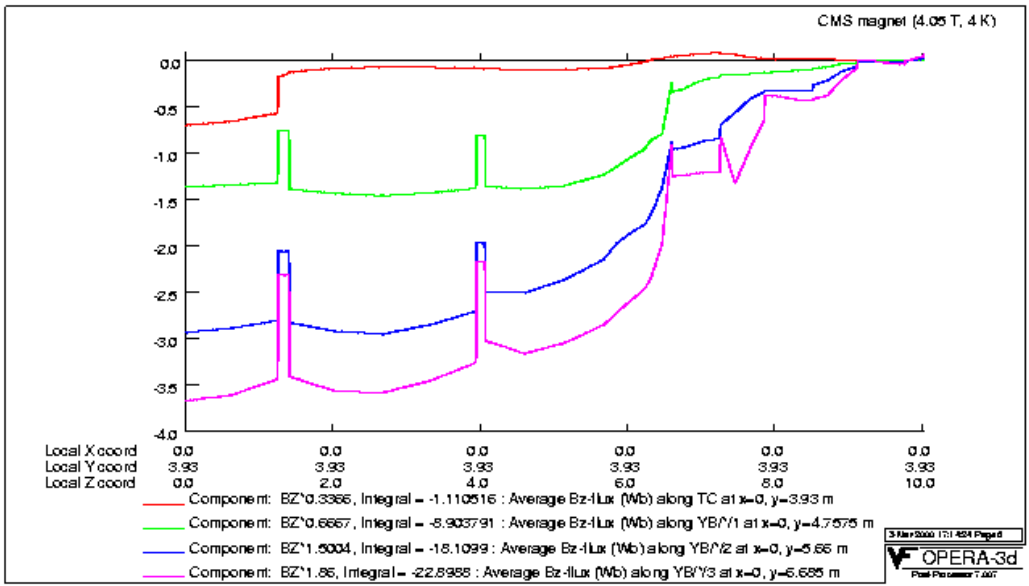

Figure 4: Average axial magnetic flux inside the barrel yoke plates.

Recalling that a flux loops measures only the magnetic flux of the field component orthogonal to the loop cross-section, we can see for example in Fig. 5 that we will not measure a substantial part of the total flux because the loops are insensitive to the axial component of the magnetic flux density.

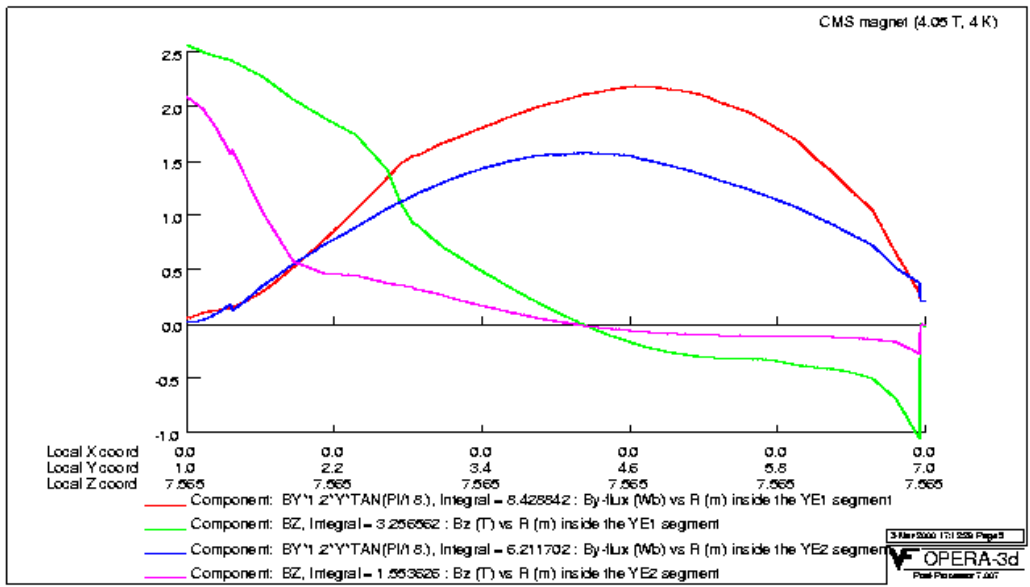

Figure 5: Average radial magnetic flux and average axial magnetic flux density inside the end-cap segments.

In summary, the flux loops around the steel plates will measure the average axial flux densities in the barrel plates presented in Fig. 6, and the average radial flux densities in the end-cap disks presented in Fig. 7.

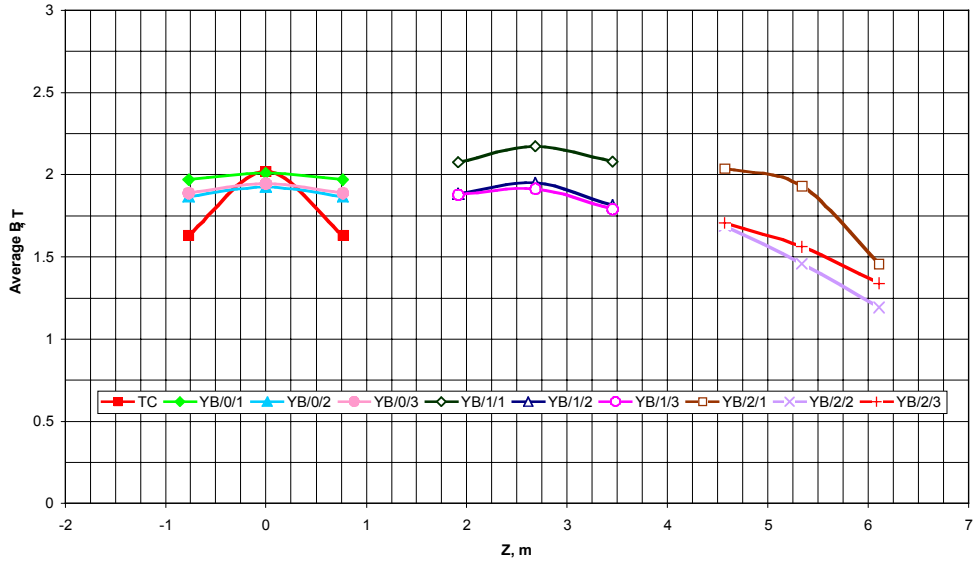

Figure 6: Average axial magnetic flux densities inside the barrel yoke plates.

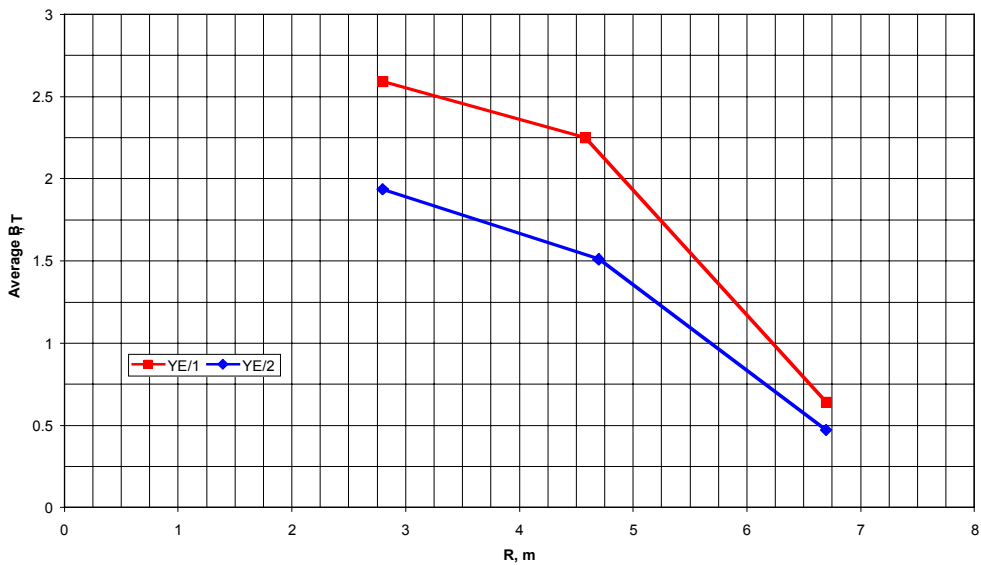

Figure 7: Average radial magnetic flux densities inside the end-cap yoke segments.

## 4  Results and discussion

The decrease of the magnetic flux during solenoid fast discharge will generate in the wire loops around the yoke plates a time dependent EMF. In each time interval $\Delta t$ after the beginning of the discharge we can determine the average induced voltage per single-turn flux loop as $V = \Delta \Phi /\Delta t$, where $\Delta \Phi$ is an amount of the flux change in the loop during the time $\Delta t$. Using eight time intervals from 0 to 305.881 s we have calculated the EMF per single-turn loop in the proposed flux loops around the plates of the tail catcher, the YB/1, YB/2, and YB/3 rings and in the proposed flux loops around the sectors of the YE/1 and YE/2 disks. These results are presented in Figs. 8-11.

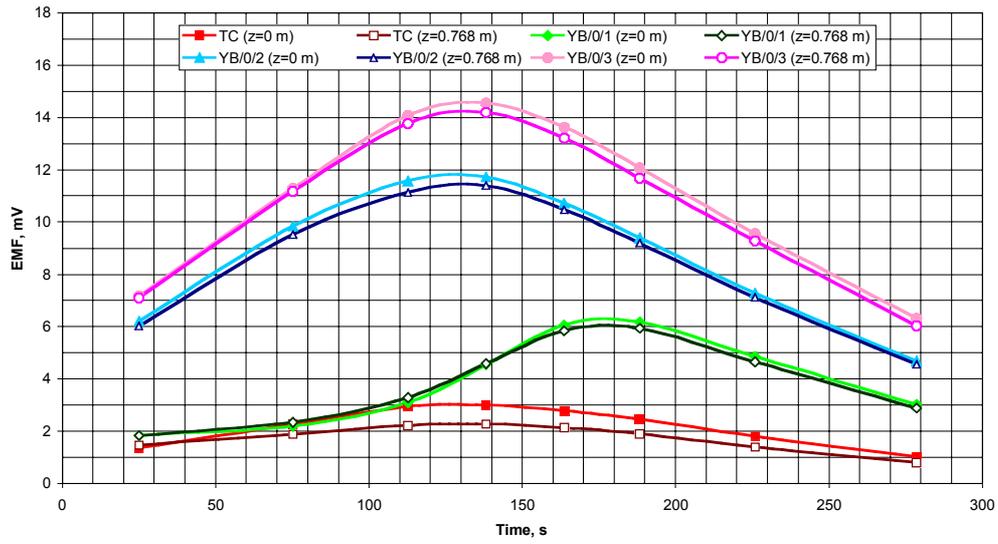

Figure 8: EMF per single-turn flux loop in the TC and YB/0 ring during the solenoid fast discharge.

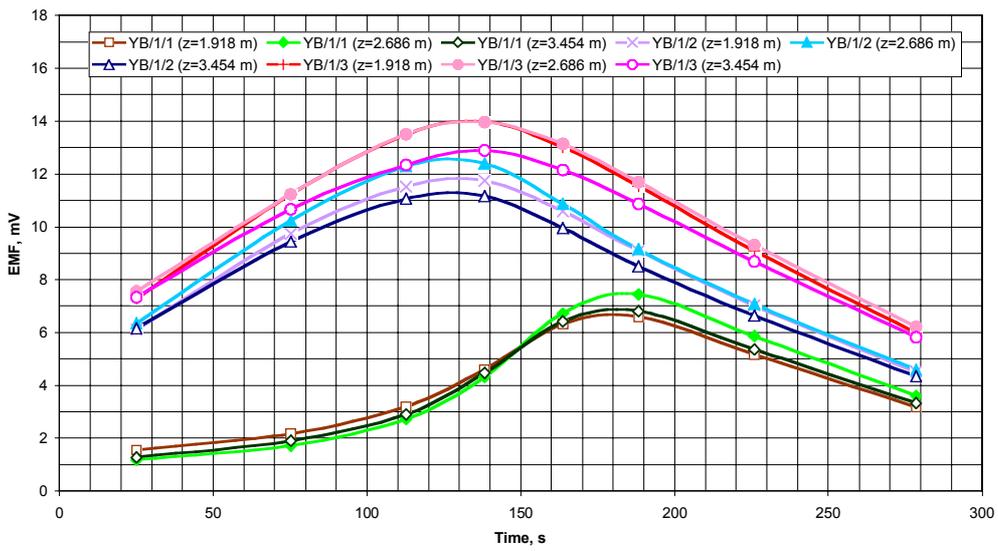

Figure 9: EMF per single-turn flux loop in the YB/1 ring during the solenoid fast discharge.

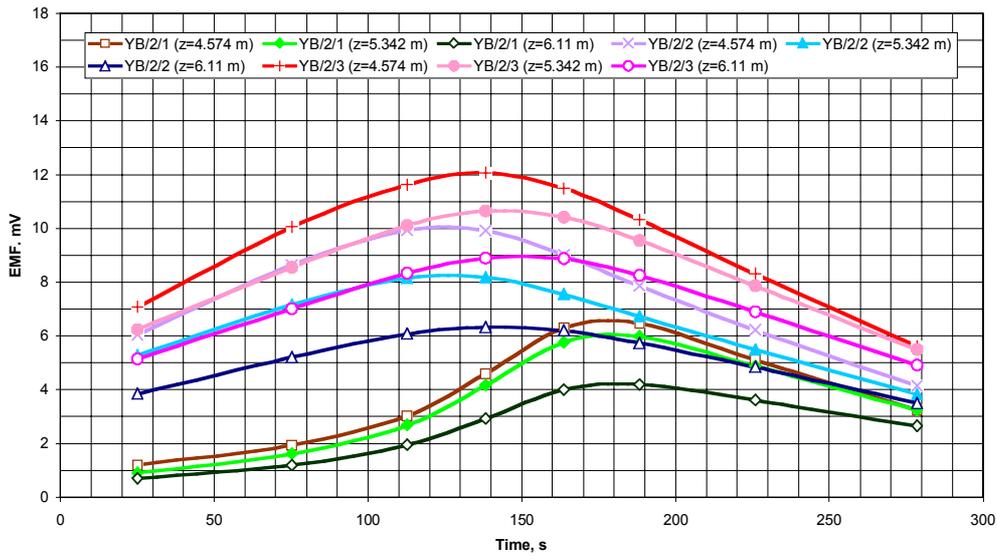

Figure 10: EMF per one wire loop turn in the YB/2 ring during the solenoid fast discharge.

In the tail catcher and barrel ring plates the EMF's reach maxima of 2 – 14 mV in 130 – 180 s after the beginning of the solenoid fast discharge and then decreases smoothly thereafter. It is seen that even at 300 seconds (when the solenoid is approximately 1/3 discharged) there is still significant voltage on each loop. Because the flux is obtained by integrating the curves from $t = 0$ to infinity, if the observations are truncated at any finite time then the unobserved portion of the discharge will constitute one source of uncertainty in the proposed procedure.

In the end-cap disks the situation is more complicated. Note in the YE/2 disk the EMF reaches the maxima of 3 – 8 mV in 140 – 160 s after the beginning of the solenoid fast discharge and then decreases, and in the YE/1 disk the EMF reaches the same maxima in 230 – 300 s from the start of the fast discharge. The rise in the EMF in YE/1 is delayed because of the heavy saturation of the iron in this disc.

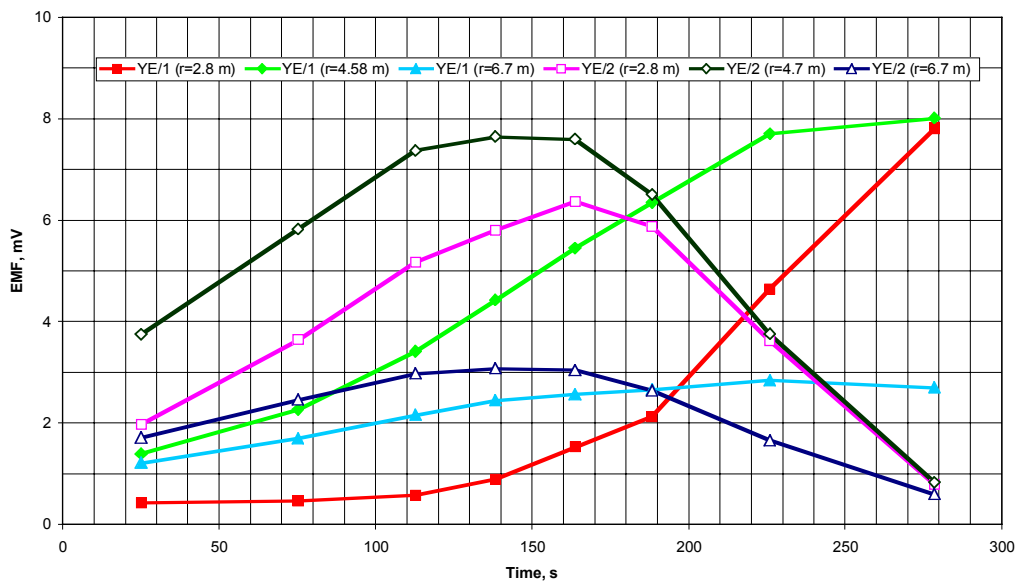

Figure 11: EMF per single-turn flux loop in the YE/1 and YE/2 disks during solenoid fast discharge.

## 4  Conclusions

These calculations indicate that a possibility to measure the EMF induced by the magnetic flux variation during the solenoid fast discharge exists.   Because of the long integration times contemplated (~ 1000 seconds), it seems preferable to sample the voltages at some moderate frequency, digitize them, and then perform the integrations numerically.

The number of turns to be provided for each flux loop can be selected so that the voltage signal is made sufficiently large to integrate meaningfully for as long as possible during the discharge.   The potential presence of electrical noise and its influence on the signals from the flux loops must also be considered.   Experience has generally shown that typically noise "integrates out" of a slowly varying signal such as that expected from the flux loops.

Depending on the magnitude of the zero-field remanence of the iron elements in the yoke, the measurements of the flux changes by the proposed technique during fast discharge of the magnet can be related to the corresponding flux changes generated by energization of the magnet.

## References

[1]  **CERN/LHCC 97-10, CMS TDR 1, May 2, 1997**, CMS, *"The Magnet Project. Technical Design Report"*.

[2]  **DSM/DAPNIA/STCM Technical Report 5C 2100T – 1000 032 98, November 18, 1998**, B. Curé, C. Lesmond, *"Synthesis on Fast Discharge Studies"*.

[3]  Vector Fields Limited, 24 Bankside, Kidlington, Oxford OX5 1JE, England.

[4]  **CMS Note 1998/048, July 30, 1998**, F. Kircher, V. Klioukhine, B. Levesy et al., *"A Review of the Magnetic Forces in the CMS Magnet Yoke"*.

[5]  DSM/DAPNIA/STCM Technical Report 5C 2100T – 0000 057 PE, October 19, 1999, B. Levesy, *"CMS Coil Parameter Book"*.